\documentclass{article}
\usepackage{spconf}
\usepackage{cite}
\usepackage{amsmath,amssymb,amsfonts}
\usepackage{algorithmic}
\usepackage{graphicx}
\usepackage{textcomp}
\usepackage{xcolor}
\usepackage{amsfonts}
\usepackage{algorithmic}
\usepackage{algorithm}
\usepackage{algorithmic}
\usepackage{hyperref}
\usepackage{subfigure}
\usepackage{tikz}
\usepackage{enumitem}
\usepackage{array, makecell}
\usepackage{multirow}
\usepackage{hhline}
\usepackage{xcolor,colortbl}
\usepackage{cite}
\usepackage{algorithm}
\usepackage{algorithmic}
\usepackage{enumitem}
\usepackage{booktabs} 
\usepackage{dcolumn} 
\usepackage{threeparttable} 
\usepackage{adjustbox} 
\usepackage{tabularx}

\newcolumntype{L}{>{\RaggedRight\arraybackslash}X}
\begin{document}
	
	\title{Semi Supervised Learning for Few-shot Audio Classification by Episodic Triplet Mining}
	\name{Swapnil Bhosale, Rupayan Chakraborty, Sunil Kumar Kopparapu}
	\address{TCS Research and Innovation -- Mumbai, INDIA
		\\
		\small{email:\{bhosale.swapnil2, rupayan.chakraborty, sunilkumar.kopparapu\}@tcs.com}}
	

\ninept
\maketitle
\begin{abstract}
	Few-shot learning aims to generalize unseen classes that appear during testing but are unavailable during training. Prototypical networks incorporate few-shot metric learning, by constructing a class prototype in the form of a mean vector of the embedded support points within a class. The performance of prototypical networks in extreme few-shot scenarios (like one-shot) degrades drastically, mainly due to the desuetude of variations within the clusters while constructing prototypes. In this paper, we propose to replace the typical prototypical loss function with an Episodic Triplet Mining (ETM) technique. The conventional triplet selection leads to overfitting, because of all possible combinations being used during training. We incorporate episodic training for mining the semi hard positive and the semi hard negative triplets to overcome the overfitting. We also propose an adaptation to make use of unlabeled training samples for better modeling. Experimenting on two different audio processing tasks, namely speaker recognition and audio event detection; show improved performances and hence the efficacy of ETM over the prototypical loss function and other meta-learning frameworks. Further, we show improved performances when unlabeled training data are used.
\end{abstract}

\section{Introduction}
\label{sec:intro}

Supervised deep learning models rely heavily on the availability of substantial amount of labeled training data. 
However, in any classification task such as identification of rare objects (unique species of birds) \cite{chencloser}, diagnosing uncommon diseases \cite{lidifficulty}, authenticating a new employee in large enterprise (i.e. speech biometry) \cite{baldwinbeyond}, detection of rare acoustic events (e.g. in audio surveillance) \cite{Zhang2020MTFCRNNMT,8673582,Park2019SoundLE}, it is a challenge to create generalizable models using traditional deep neural networks. 
On the other hand, humans utilize their past experience in order to learn new concepts across different domains. And conversely, most deep learning models are able to learn high level features and extract complex characteristics, only when sufficient amount of labeled data is used for supervised training \cite{najafabadideep}.

Few-shot learning has the potential to create generalizable models as it re-frames the learning paradigm such that the model is not trained to classify a sample into one of the categories seen during training. But it is optimized such that, given a pair of samples, it can predict whether the two samples are similar or dissimilar \cite{NIPS2017_6820}\cite{sung2018learning}\cite{snellprototypical}. 

An extreme case of few-shot learning is one-shot learning, where a single sample (i.e. a reference) for each of the unseen classes is available for making an inference about a test sample.
Many few-shot learning problems utilize meta learning algorithms that learn a mapping to an embedding space, where samples belonging to the same classes are closer, compared to those belonging to different classes \cite{sanakoyeudivide}. One such popular framework is Prototypical networks that learns an embedding space, where the samples within the same class form clusters around a single prototypical point, which is represented by the mean of individual samples within the cluster. During inference, a query sample is assigned the label corresponding to the nearest prototype in the embedding space. There are two prominent problems with such approach. First, variance in the data can easily affect the relative positions of the prototypes since the model relies on the unweighted average of the samples. Second, for extreme cases like one-shot, it assumes each individual sample to be a separate cluster, with sample itself being the prototype. Hence, the performance of prototypical networks trained on multiple shots, degrades drastically in one-shot inference setup.

To address these problems, we propose to use the episodic triplet loss instead of the typical prototypical loss function. In addition, conventional triplet selection during the training process is cumbersome since choosing all possible selections, can easily lead to over-fitting. Therefore, we incorporate episodic training for mining the semi hard positive and the semi hard negative triplets. To validate our proposal, we conducted experiments over two different speech processing tasks, namely, (1) Speaker recognition task (using VCTK corpus), (2) Audio Event Classification task (using Freesound dataset, 2018). Our results show improved performances, especially in case of extreme few shot test setups, and indicate the efficacy of using the episodic triplet loss over the prototypical loss function. 
To summarize, the main contributions of this paper are, (a) A novel few-shot learning approach by replacing the conventional prototypical loss function with a Episodic Triplet Mining technique (ETM). This is more useful in case of extreme few-shot scenario (e.g. one-shot learning). (b) Incorporate episodic training for mining the semi hard positive and the semi hard negative triplets, particularly to avoid the over-fitting that arises due to the usual all possible triplet mining strategy. (c) An adaptation of our proposed approach to effectively use the unlabeled samples available during training in a semi supervised paradigm. We experimentally validate the proposed approach in semi supervised scenarios, and show that our model trained using a subset of labeled training data performs competitively with the models that are trained in supervised manner on complete set of labeled data. The rest of the paper is organized as follows. In Section \ref{sec:prior}, we provide a brief literature review of the related work in this area. In Section \ref{sec:proposed}, we explain in detail the system design and the proposed approach. The experimental details, and results are presented in Section \ref{sec:exp}, followed by conclusion in Section \ref{sec:conclusion}.

\section{Related Work}
\label{sec:prior}
Prototypical networks \cite{snellprototypical} and Matching networks \cite{vinyalsmatching} are the most popular metric learning based methods for few-shot learning. In prototypical networks, there exists an embedding space in which samples belonging to the same class form distinct clusters. During the training process, prototypes for each class are computed using the average of the embeddings of all samples having that class as their label. That is why the prototypes do not effectively capture the variations in data. Moreover, extensions of prototypical loss in extreme few-shot cases, such as one-shot, assume each point to be an individual cluster with its embedding as the prototype. \cite{wang2020few}, \cite{shi2020few} adapted prototypical networks for AED and showed generalization to unseen audio events during real-time. In \cite{anandfew}, authors proposed a prototypical loss based few-shot learning architecture for speaker recognition, where capsule networks are used to extract audio embeddings from input Mel-spectrograms features. In contrast, we use a simpler embedding network with 1D convolutional layers, with self-attention on the outputs of convolution layers, and the average of attention heads is used as the embedding for input spectrogram. Authors in \cite{wangcentroid}, compare the triplet loss with prototypical loss for speaker recognition task. Although their results favor prototypical loss over triplet loss, but we hypothesize that the triplet mining strategy could severely affect the performance of such networks. In traditional triplet frameworks, given a triplet, the interaction of the anchor point is limited to a single positive and negative point. In order to optimize the anchor embedding with respect to different points, the same point must have multiple entries as anchor in various triplets. As a result,  the number of possible training triplets can grow rapidly with the size of the dataset, thus making it impractical. In this direction, we propose an episodic training framework for triplet loss, wherein each episode optimizes the triplet loss with respect to each query point (as an anchor). By doing so, in each episode a single anchor point now simultaneously interacts with all the positive and negative samples present in the support set.

Recently in \cite{boneysemi}, \cite{ayyadsemi} and \cite{lilearning}, authors have extended prototypical networks for incorporating unlabeled data samples, by adapting a semi supervised paradigm. To amplify the efficacy of triplet loss trained with episodic framework, we adapt our approach to semi supervised paradigms as well. During the training process, we first incorporate pseudo-labeling for the unlabeled samples in each episode, and then combine it with the existing support set. The query points are now optimized by comparing the distances of positive and negative samples from the new support set. Through extensive experimentation, we empirically show the influence of amount of unlabeled samples available in each episode on the performance of the system.


\section{System Design}

\label{sec:proposed}

Consider a $N$-shot $K$-way learning problem such that $N$ samples from each of the $K$ unique classes are provided. Each sample is represented by a $F$-dimensional feature vector $x_{i} \in \mathcal{R}^F$, and its label $y_{i} \in \{1, ..., K\}$. A Support set is defined as $\big\{{x_{i},y_{i}}\big\}_{i=1}^{N_{S}}$, 
where $N_{S}$ is the number of support samples. Similarly, a Query set is defined as 
$\big\{{x_{i}}\big\}_{i=1}^{N_{Q}}$, 
where $N_{Q}$ is the number of queries. The embedding function $f_{\phi}(.)$, parameterized by weights $\phi$, projects each sample $x_{i}$ to an $M$ dimensional embedding in the latent space. 
$f_{\phi}(.)$ is optimized to reduce the distance between the embeddings of queries and support samples belonging to the same class and increase the distance between queries and support samples belonging to different classes.


\begin{figure}[!t]
	\centering
	\includegraphics[width=0.9\linewidth]{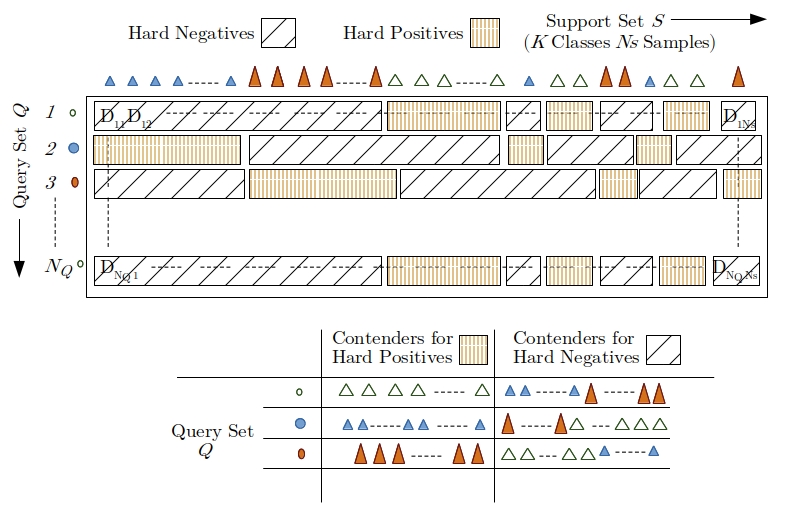}
	\caption{Selection of hard positives and hard negatives from distance matrix, $D$, for a single episode.}
	\label{fig:matrix}
\end{figure}

\subsection{Proposed Approach}

$E_{S}$ and $E_{Q}$ denote the embeddings for all samples within the support set and query set respectively. A distance matrix $D$ is formed with each embedding in $E_{Q}$ and $E_{S}$ placed along the rows and columns respectively, with value at each cell as,
$D_{i,j} = d_{e}(E_{Q}[i], E_{S}[j]),$
where $i = \{1, ..., N_{Q}\}, j = \{1, ..., N_{S}\}$ and $d_{e}(.)$ is the euclidean distance. For each query $q_{i} \in E_{Q}$, the samples within the support set, belonging to the same class as that of $q_{i}$, constitute the set of positive samples and all remaining samples from the support, form a set of negative samples (see Figure \ref{fig:matrix}).

Triplet loss function compares the distance of an input sample (anchor), to its distance from a positive input (same class as anchor) and a negative input (different class than anchor). Within a training episode, the triplet loss is calculated for each query sample $q_{i}$ treated as the anchor and $f_\phi(.)$ is optimized such that, $(d^{N}_{q_{i}} - d^{P}_{q_{i}}) > z$, 
where $d^{N}_{q_{i}}$ is the distance between $q_{i}$ and the negative set, $d^{P}_{q_{i}}$ is the distance between $q_{i}$ and the positive set, and $z$ is the margin. Considering $q_{i} \in E_{Q}$, $d^{P}_{q_{i}}$ and $d^{N}_{q_{i}}$ can be represented as follows,
\begin{equation*}
\label{eqn:2}
d^{P}_{q_{i}} = \mu(\Lambda_{max}(d_{e}(q_{i}, s_{j}), n_P)),
\forall s_{j} \in E_{S} \mid l(s_{j}) = l(q_{i})
\end{equation*}
\begin{equation*}
\label{eqn:3}
d^{N}_{q_{i}} = \mu(\Lambda_{min}(d_{e}(q_{i}, s_{j}), n_N)),
\forall s_{j} \in E_{S} \mid l(s_{j}) \neq l(q_{i})
\end{equation*}
where $\mu$ is the mean operator, $\Lambda_{max}(A,n)$ samples highest $n$ values from set $A$, $\Lambda_{min}(A,n)$ samples lowest $n$ values from set $A$ and $l(.)$ returns the corresponding class label. 
Instead of choosing only the farthest positive and the nearest negative sample, we average the top $n_{P}$ samples from the positive set which are farthest from $q_{i}$ and average the top $n_{N}$ samples from the negative set which are nearest from $q_{i}$. 

In traditional triplet loss \cite{schroff2015facenet}, the number of possible triplets to be passed through $f_\phi(.)$ grows quadratically with number of samples due to which the same anchor point gets repeated in multiple triplets, thus leading to overfitting. Conversely, in ETM, each anchor simultaneously interacts with all negative samples in the episode and hence, gives a more stable update and converges faster. The $n_N$ parameter in ETM, chooses semi hard negatives among hard negatives of individual classes in each episode.
Algorithm \ref{alg:fsl} explains the procedure for computing the loss for a single episode in detail. Similar to the training phase, during inference for each query sample, we identify top $n_{P}$ nearest samples from the support set, and assign the label accordingly.

%
%
\begin{algorithm}[tb]
	\caption{Triplet loss for a single episode.}
	\label{alg:fsl}
	\textbf{Input}: Train set, $X = \{(x_{1},y_{1}), ..., (x_{N},y_{N})\}$, where $y_{i}\in\{1, ..., K\}$.\\
	$X_{k}$ denotes the subset of $X$ containing all elements ($x_{i}$,$y_{i}$) such that $y_{i} = k$. \\
	$N_{c}$ : number of unique classes in an episode.\\
	$N_{S}$ : number of support samples for each class.\\
	$N_{Q}$ : number of query samples for each class.\\
	$\Lambda(X,n)$ : randomly samples a subset of size $n$ from set $X$\\
	\textbf{Output}: triplet loss for each episode.\\
	\begin{algorithmic}[1] 
		\STATE $V \leftarrow \Lambda(\{1,...,K\},N_{c}).$
		\FOR{$k$ in \{1, ..., $N_{c}$\}}
		\STATE $S_{k} \leftarrow \Lambda(X_{V_{k}}, N_{S})$
		\STATE $Q_{k} \leftarrow \Lambda(X_{V_{k}}\setminus{S_{k}}, N_{Q})$
		\ENDFOR
		\STATE $S \leftarrow \{S_{1}, ..., S_{N_{c}}\}$
		\STATE $Q \leftarrow \{Q_{1}, ..., Q_{N_{c}}\}$
		\STATE $E_{S} \leftarrow f_{\phi}(x_{i}) \forall x_{i} \in S$
		\STATE $E_{Q} \leftarrow f_{\phi}(x_{i}) \forall x_{i} \in Q$
		
		\STATE $D_{ij} \leftarrow d_{e}(E_{Q}[i], E_{S}[j]) \forall i \in \{1, ..., N_{Q}\} \forall j \in \{1, ..., N_{S}\}$
		
		\FOR{$q_{i} \in E_{Q}$}
		
		\STATE $loss_{i} = \max{(d^{P}_{q_{i}} - d^{N}_{q_{i}} + z, 0)}$
		\ENDFOR
		\STATE $loss \leftarrow \sum loss_{i}$
	\end{algorithmic}
	
\end{algorithm}
\begin{figure}[!t]
	\centering
	\includegraphics[width=0.8\linewidth]{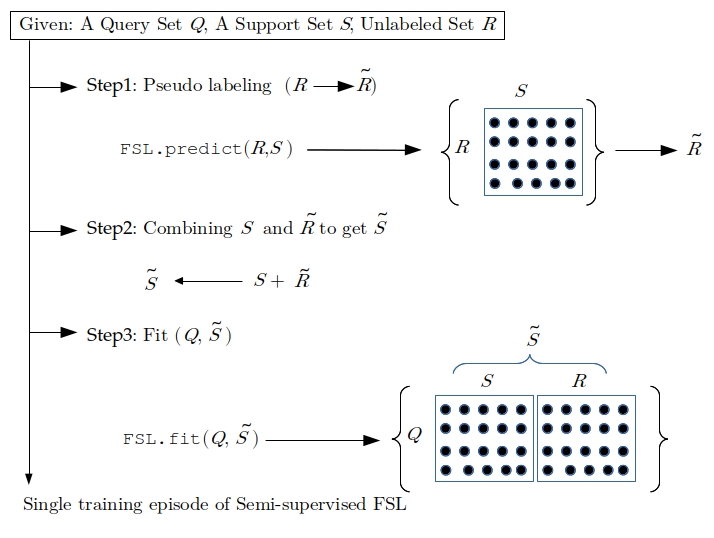}
	\caption{Single training episode for semi supervised few-shot learning.}
	\label{fig:ssl}
\end{figure}

\subsubsection{Adaptation to semi supervised learning}

In a semi supervised system, along with a support set $S$ and a query set $Q$, we are also provided with an unlabeled set, $R$. The unlabeled samples can be used for regularization, and to overcome the overfitting when labeled training samples are scarce. We adopt a pseudo-labeling technique wherein each training episode, we first infer the labels for each sample in $R$ (considering it as a query) based on labels of the nearest samples within $S$ and denote it as $\tilde{R}$. Next, we combine $S$ and $\tilde{R}$ to form our new support set, $\tilde{S}$, which is used to calculate the loss over each query in $Q$. We hypothesize that the pseudo-labeling in the first step of every episode induces an error in the system. This gets further propagated while optimizing the triplet loss over query samples with $\tilde{S}$ as support set. As a result, the overfitting is delayed since there is always some loss that gets backpropagated through the embedding network. Figure \ref{fig:ssl} depicts the steps in a single training episode for semi-supervised learning.

We choose two different scenarios, (a) weakly labeled (b) completely unlabeled. In weakly labeled setup, for a particular episode, the unlabeled samples $R$ are chosen from the same classes that are present in the $S$. On the other hand, the completely unlabeled setup resembles a more realistic scenario, in which no metadata is involved while choosing the samples in $R$. The classes in $R$ may be either present within $S$ or maybe completely disjoint. In contrast to \cite{renmeta}, we refrain from using the distractor classes, since choosing distractor samples requires prior information about labels, which may not be always available.

\section{Experiments}
\vspace{-0.2cm}
\label{sec:exp}
To validate our proposed approach, experiments are conducted for two different tasks, namely, (1) Speaker Recognition (VCTK corpus \cite{veauxsuperseded}) (2) Audio Event Classification (Freesound Dataset, 2018 (FSD) \cite{fonseca2018general}).

\renewcommand{\paragraph}[1]{{\vspace{\baselineskip}\noindent\normalfont\bfseries#1\quad}}
\paragraph{Speaker Recognition task :}
VCTK corpus is an English multi-speaker dataset, with 44 hours of audio spoken by 109 native English speakers. We split the dataset into 70:20:10 random train-test-validation split, such that set of speakers in train, test and validation set are completely disjoint. We down-sampled each audio to 16 kHz, and split into audio segments of 3 seconds each. Mel-spectrograms are extracted as an initial feature from each segment and used as an input to the embedding network. The embedding network is constructed using two layers of 1-D convolutions each with a kernel of size 3 and with 128 filters each. The use of 1-D convolution helps learn the temporal contexts between the adjacent frames. Each convolution layer is followed by a max-pooling layer with kernel of size 3. Additionally, batch normalization is performed over the output of each convolution layer. We apply a multi-head self attention mechanism \cite{vaswani2017attention} over the output of the second convolution layer and average the output of each head to obtain an 128 dimensional embedding.

The semi supervised learning experiments are conducted considering the availability of two different portions of the training data as labeled train set, (a) 33\% of the samples of each speaker present in the training data and (b) 66\% of the samples of each speaker present in the training data. In both (a) and (b), the remaining samples from the training data are used as the unlabeled train set.

\paragraph{Audio Event Classification task :}
Freesound dataset (2018) consists of 18,873 audio files wherein each audio file is assigned one of the 41 unique audio events from the Google's Audioset Ontology \cite{gemmeke2017audio}. In each of the 3 folds of our experiment, we randomly choose 10 classes with all its corresponding audio files as our test set, and split remaining classes into train classes and validation classes in 90:10 ratio, such that the audio events in train, test and validation are always disjoint. All audio files are down-sampled to 16 kHz, and split into 1 second chunks. We use a VGGish architecture \cite{hershey2017cnn} as our embedding model, which outputs a 128 dimensional embedding for each chunk of audio. For the semi supervised learning experiments, we randomly choose 50\% of samples from each audio event within the training data, and mark it as our labeled training set. The rest of the samples forms our unlabeled train set. The test set and validation set for the supervised and semi supervised experiments for each task are kept fixed.

\paragraph{Episode construction}

For speaker recognition we found $N_{S}$ = 20, $N_{Q}$ = 15 and $N_{c}$ = 5, to give the best performance across all testing configurations. Similarly, for the audio event classification task, we found $N_{S}$ = 10, $N_{Q}$ = 5 and $N_{c}$ = 5, to give the best results. In all tasks, we choose $z$ (i.e. margin) as 0.3. Also, we train a single model, and test it against all test configurations (5-way 1-shot, 5-way 5-shot and so on). We use $n_{P}$ = 3 and $n_{N}$ = 5 for all tasks. For speaker recognition and audio event classification task, we train our model for 10,000 episodes. We use the Adam optimizer with initial learning rate of $10^{-3}$ which gets reduced by half after every 1000 episodes. While testing, we use $N_{Q}$ = 15, and report the average accuracy over 1000 test episodes.

For the semi supervised setup, we found that pre-training the model for a few initial episodes (50 episodes for both speaker recognition and audio event classification) in a completely supervised setup results in slightly smoother training curves. This might be due to the fact that the incorporating the unlabeled data from the first episode itself might result in a higher value of error, thus making the convergence difficult.
We vary the amount of unlabeled samples per class in each episode from 1 to 5. We employ a similar testing setup as the supervised setup, keeping the test set unchanged. 

\subsection{Results and Analysis}
\vspace{-0.2cm}
Tables \ref{tab:spk_1}-\ref{tab:spk_3} and Tables \ref{tab:aed_1}-\ref{tab:aed_2} show the results for speaker recognition and audio event classification tasks, respectively. We evaluate the performance of all the experiments in terms of average accuracy across all test episodes. 

Table \ref{tab:spk_1} shows results for few-shot speaker recognition on the entire training data. We compare our approach with prototypical loss, matching network, relation network all with the same embedding module. We surpass the accuracy of the model trained using prototypical loss in \textit{5-way 1-shot}, \textit{20-way 1-shot}, \textit{20-way 5-shot}. Especially in \textit{20-way 1-shot} and  \textit{5-way 1-shot}, our approach outperforms a model trained using prototypical loss with identical embedding network by an absolute of 11.17\% and 3.24\%, respectively. 

\begin{table}[]
\caption{Performance comparison for speaker recognition task. (entire train data is labeled)}
\label{tab:spk_1}
\begin{adjustbox}{max width=\linewidth}
\centering

\begin{tabular}{|c|c|c|c|c|}
\hline
loss function                                                                         & \begin{tabular}[c]{@{}c@{}}5-way 1-shot\end{tabular} & \begin{tabular}[c]{@{}c@{}}5-way 5-shot\end{tabular} & \begin{tabular}[c]{@{}c@{}}20-way 1-shot\end{tabular} & \begin{tabular}[c]{@{}c@{}}20-way 5-shot\end{tabular} \\ \hline 
\hline
\begin{tabular}[c]{@{}c@{}}Matching Network \end{tabular}                      &                              82.77                       &   93.79                                                  &               55.02                                        &         76.29                                               \\
\hline
\begin{tabular}[c]{@{}c@{}}Relation Network \end{tabular}                      & 82.68                                                    & 94.31                                                    & 54.55                                                      & 76.12                                                       \\
\hline
\begin{tabular}[c]{@{}c@{}}Prototypical loss \end{tabular}                      & 83.42                                                    & \bf{96.88}                                                     & 56.23                                                      & 78.47                                                       \\ \hline
\begin{tabular}[c]{@{}c@{}}Episodic Triplet Mining (ETM)\end{tabular} & \bf{86.66}                                                    & 93.04                                                     & \bf{67.4}                                                       & \bf{79.15}                                                       \\ \hline
\end{tabular}
\end{adjustbox}
\vspace{-0.3cm}
\end{table}

\begin{table}[]
\caption{Speaker recognition performance of semi supervised learning system using weakly labeled and completely unlabeled data. (only 33\% of the training data is labeled)}
\label{tab:spk_2}
\begin{adjustbox}{max width=\linewidth}
\centering
\begin{tabular}{|c|c|c|c|c|}

\hline

Type                                                                         & \begin{tabular}[c]{@{}c@{}}5-way 1-shot\end{tabular} & \begin{tabular}[c]{@{}c@{}}5-way 5-shot\end{tabular} & \begin{tabular}[c]{@{}c@{}}20-way 1-shot\end{tabular} & \begin{tabular}[c]{@{}c@{}}20-way 5-shot\end{tabular} \\ \hline \hline

Supervised (baseline)                                                        & 76.13                                                  & 87.90                                                  & 49.66                                                   & 65.45                                                    \\ \hline

\begin{tabular}[c]{@{}c@{}}Semi supervised (Weakly labeled)\end{tabular}  & \textbf{85.58}                                         & 91.37                                                  & \textbf{64.44}                                          & \textbf{76.90}                                           \\ \hline

\begin{tabular}[c]{@{}c@{}}Semi supervised (completely unlabeled)\end{tabular} & 85.41                                                  & \textbf{92.13}                                         & 63.65                                                   & 75.39                                                    \\ \hline \hline

\begin{tabular}[c]{@{}c@{}}Top-line (100\% labeled data)\end{tabular}         & 86.66                                                  & 93.04                                                  & 67.40                                                   & 79.15                                                    \\ \hline

\end{tabular}
\end{adjustbox}

\vspace{-0.3cm}
\end{table}

\begin{table}[!t]
\centering

\caption{Speaker recognition performance of semi supervised learning system using weakly labeled and completely unlabeled data. (only 66\% of the training data is labeled)}

\label{tab:spk_3}
\begin{adjustbox}{max width=\linewidth}
\begin{tabular}{|c|c|c|c|c|}

\hline

Type                                                                         & \begin{tabular}[c]{@{}c@{}}5-way 1-shot\end{tabular} & \begin{tabular}[c]{@{}c@{}}5-way 5-shot\end{tabular} & \begin{tabular}[c]{@{}c@{}}20-way 1-shot\end{tabular} & \begin{tabular}[c]{@{}c@{}}20-way 5-shot\end{tabular} \\ \hline \hline

Supervised (baseline)                                                        & 84.12                                                  & 92.14                                                  & 63.53                                                   & 76.32                                                    \\ \hline

\begin{tabular}[c]{@{}c@{}}Semi supervised (Weakly labeled)\end{tabular}  & 85.28                                                  & 92.34                                         & \textbf{65.10}                                                   & 77.59                                                    \\ \hline

\begin{tabular}[c]{@{}c@{}}Semi supervised (completely unlabeled)\end{tabular} & \textbf{87.05}                                         & \textbf{92.84}                                                  & 65.00                                        & \textbf{78.37}                                           \\ \hline \hline

\begin{tabular}[c]{@{}c@{}}Top-line (100\% labeled data)\end{tabular}         & 86.66                                                  & 93.04                                                  & 67.40                                                   & 79.15                                                    \\ \hline

\end{tabular}
\end{adjustbox}

\vspace{-0.3cm}
\end{table}

%
%
%
%
%
%

In case of semi supervised few-shot learning, while using only 33\% of the training data as labeled (see Table \ref{tab:spk_2}), our model achieves significantly better results over the baseline (which undergoes supervised training using 33\% of training data), thereby reducing the gap with respect to the top-line across all four setups by an average of 11.45\%. Also, while using 66\% of the training data as labeled (see Table \ref{tab:spk_3}), we achieve an average increase of 1.66\% across all four setups. Specifically for \textit{5-way 1-shot}, our approach achieves an absolute improvement of 0.39\% over a model trained in supervised manner using 100\% labeled training data. The psuedo labeling technique is an iterative process which instead of diffusing the probability spread over multiple classes, assigns high probability towards one particular class  \cite{lee2013pseudo}. This helps by reducing the density (or entropy \cite{grandvalet2005semi}) around the decision boundaries. Please note that the psuedo label is assigned using a few-shot inference. 
		

Table \ref{tab:aed_1} compares the performance of episodic triplet mining with existing meta-learning frameworks for the audio event classification task on various one-shot scenarios. We achieve an average improvement of 3.72\% in terms of accuracy when training the model using ETM technique across all three one-shot scenarios. Similar to the speaker recognition task, Table \ref{tab:aed_2} shows the performance of ETM extended to semi supervised domain for the audio event classification task. For both, weakly labeled and completely labeled experiments, we use 50\% of the samples belonging to each class in the train set as the unlabeled training set and the remaining samples as the labeled training set.

\begin{table}[]

\centering

\caption{Performance comparison for Audio Event Classification task. (entire train data is labeled)}

\label{tab:aed_1}
\begin{adjustbox}{max width=0.85\linewidth}

\begin{tabular}{|c|c|c|c|}

\hline

loss function                                                                         & \begin{tabular}[c]{@{}c@{}}5-way 1-shot\end{tabular} & 

\begin{tabular}[c]{@{}c@{}}7-way 1-shot\end{tabular} & 


\begin{tabular}[c]{@{}c@{}}10-way 1-shot\end{tabular} \\ \hline 
\hline
\begin{tabular}[c]{@{}c@{}}Matching Network \end{tabular}                      &75.11                        

&72.88                           


&63.96 \\
\hline
\begin{tabular}[c]{@{}c@{}}Relation Network \end{tabular}                      &74.29                        

&73.13                           


&64.61 \\
\hline
\begin{tabular}[c]{@{}c@{}}Prototypical loss \end{tabular}                      &75.33                        

&72.95                           


&63.33 \\ \hline

\begin{tabular}[c]{@{}c@{}}Episodic Triplet Mining (ETM)\end{tabular}  &\bf{78.53}                        

&\bf{75.04}                      


&\bf{69.40} \\ \hline

\end{tabular}
\end{adjustbox}
\vspace{-0.3cm}
\end{table}

\begin{table}[]

\centering

\caption{Audio event classification performance of semi supervised learning system using weakly labeled and completely unlabeled data using ETM. (only 50\% of the training data is labeled)}

\label{tab:aed_2}
\begin{adjustbox}{max width=0.85\linewidth}

\begin{tabular}{|c|c|c|c|}

\hline

Type                                                                         & \begin{tabular}[c]{@{}c@{}}5-way 1-shot\end{tabular} & \begin{tabular}[c]{@{}c@{}}7-way 1-shot\end{tabular} & \begin{tabular}[c]{@{}c@{}}10-way 1-shot\end{tabular} \\ \hline \hline

Supervised (baseline)                                                        & 76.56    & 72.34                                                                                                   &61.44                                            \\ \hline

\begin{tabular}[c]{@{}c@{}}Semi supervised (Weakly labeled)\end{tabular}

&\bf{79.20}     &  73.34                                                                                                  &  \bf{69.24}                                         \\ \hline

\begin{tabular}[c]{@{}c@{}}Semi supervised (completely unlabeled)\end{tabular}

& 77.73    & \bf{73.66}                                                                                                   & 68.33                                           \\ \hline \hline

\begin{tabular}[c]{@{}c@{}}Top-line (100\% labeled data)\end{tabular}

&78.53     &75.04                                                                                                    &69.33                                     \\ \hline

\end{tabular}
\end{adjustbox}
\vspace{-0.3cm}
\end{table}

\section{Conclusion}
\label{sec:conclusion}
\vspace{-0.2cm}
In this paper, we propose a novel Episodic Triplet Mining (ETM) for few-shot learning and compared it with the conventional Prototypical loss models. We incorporate the episodic training for mining the semi hard positive and the semi hard negative triplets, precisely to avoid the over-fitting, which arise due to usual strategy of all possible triplet mining. Proposed ETM, importantly, in case of one-shot learning, outperformed the prototypical loss based model. Moreover, we validate the usefulness of ETM over prototypical loss, especially in presence of unlabeled samples. ETM model adapted to the semi supervised learning surpasses the baseline by a large margin in both the cases, (a) in presence of weakly labeled data (i.e. when the unlabeled data is used based on the metadata), (b) in presence of completely unlabeled data. Moreover, the performance variations with respect to the amount of unlabeled data available in each training episode are also investigated. The ability of deep learning models to generalize the new classes, without the need of re-training or fine tuning is extremely useful. In addition, there is always a constraint on how many reference samples could the user be requested to provide, which makes the generalizability in the presence of minimum reference samples much more crucial. The better performance of ETM specifically for extreme few-shot scenarios (like one-shot) are being observed in our work consistently. The ETM adaptation to semi supervised domain leverages the unlabeled data which is present in abundance. 
\bibliographystyle{IEEEbib}

\bibliography{Template}

\begin{thebibliography}{10}

\bibitem{chencloser}
Wei-Yu Chen, Yen-Cheng Liu, Zsolt Kira, Yu-Chiang~Frank Wang, and Jia-Bin
  Huang,
\newblock ``A {C}loser {L}ook at {F}ew-{S}hot {C}lassification,''
\newblock {\em arXiv preprint arXiv:1904.04232}, 2019.

\bibitem{lidifficulty}
Xiaomeng Li, Lequan Yu, Chi-Wing Fu, and Pheng-Ann Heng,
\newblock ``Difficulty-{A}ware {M}eta-{L}earning for {R}are {D}isease
  {D}iagnosis,''
\newblock {\em arXiv preprint arXiv:1907.00354}, 2019.

\bibitem{baldwinbeyond}
Jacob Baldwin, Ryan Burnham, Andrew Meyer, Robert Dora, and Robert Wright,
\newblock ``Beyond {S}peech: {G}eneralizing {D}-{V}ectors for {B}iometric
  {V}erification,''
\newblock in {\em Proceedings of the AAAI Conference on Artificial
  Intelligence}, 2019, vol.~33, pp. 842--849.

\bibitem{Zhang2020MTFCRNNMT}
Keming Zhang, Y.~Cai, Y.~Ren, Ruida Ye, and L.~He,
\newblock ``M{TF}-{CRNN}: {M}ultiscale {T}ime-{F}requency {C}onvolutional
  {R}ecurrent {N}eural {N}etwork for {S}ound {E}vent {D}etection,''
\newblock {\em IEEE Access}, vol. 8, pp. 147337--147348, 2020.

\bibitem{8673582}
A.~{Mesaros}, A.~{Diment}, B.~{Elizalde}, T.~{Heittola}, E.~{Vincent},
  B.~{Raj}, and T.~{Virtanen},
\newblock ``Sound {E}vent {D}etection in the {DCASE} 2017 {C}hallenge,''
\newblock {\em IEEE/ACM Transactions on Audio, Speech, and Language
  Processing}, vol. 27, no. 6, pp. 992--1006, 2019.

\bibitem{Park2019SoundLE}
Jeong-Sik Park and Seok-Hoon Kim,
\newblock ``Sound {L}earning based {E}vent {D}etection for {A}coustic
  {S}urveillance {S}ensors,''
\newblock {\em Multimedia Tools and Applications}, vol. 79, pp. 16127--16139,
  2019.

\bibitem{najafabadideep}
Maryam~M Najafabadi, Flavio Villanustre, Taghi~M Khoshgoftaar, Naeem Seliya,
  Randall Wald, and Edin Muharemagic,
\newblock ``Deep {L}earning {A}pplications and {C}hallenges in {B}ig {D}ata
  {A}nalytics,''
\newblock {\em Journal of Big Data}, vol. 2, no. 1, pp. 1, 2015.

\bibitem{NIPS2017_6820}
Eleni Triantafillou, Richard Zemel, and Raquel Urtasun,
\newblock ``Few-{S}hot {L}earning {T}hrough an {I}nformation {R}etrieval
  {L}ens,''
\newblock in {\em Advances in Neural Information Processing Systems}, pp.
  2255--2265. Curran Associates, Inc., 2017.

\bibitem{sung2018learning}
Flood Sung, Yongxin Yang, Li~Zhang, Tao Xiang, Philip~HS Torr, and Timothy~M
  Hospedales,
\newblock ``Learning to {C}ompare: {R}elation {N}etwork for {F}ew-{S}hot
  {L}earning,''
\newblock in {\em Proceedings of the IEEE Conference on Computer Vision and
  Pattern Recognition}, 2018, pp. 1199--1208.

\bibitem{snellprototypical}
Jake Snell, Kevin Swersky, and Richard Zemel,
\newblock ``Prototypical {N}etworks for {F}ew-{S}hot {L}earning,''
\newblock in {\em Advances in Neural Information Processing Systems}, 2017, pp.
  4077--4087.

\bibitem{sanakoyeudivide}
Artsiom Sanakoyeu, Vadim Tschernezki, Uta Buchler, and Bjorn Ommer,
\newblock ``Divide and {C}onquer the {E}mbedding {S}pace for {M}etric
  {L}earning,''
\newblock in {\em Proceedings of the IEEE Conference on Computer Vision and
  Pattern Recognition}, 2019, pp. 471--480.

\bibitem{vinyalsmatching}
Oriol Vinyals, Charles Blundell, Timothy Lillicrap, Daan Wierstra, et~al.,
\newblock ``Matching {N}etworks for {O}ne {S}hot {L}earning,''
\newblock in {\em Advances in Neural Information Processing Systems}, 2016, pp.
  3630--3638.

\bibitem{wang2020few}
Yu~Wang, Justin Salamon, Nicholas~J Bryan, and Juan~Pablo Bello,
\newblock ``Few-{S}hot {S}ound {E}vent {D}etection,''
\newblock in {\em IEEE International Conference on Acoustics, Speech and Signal
  Processing (ICASSP)}. IEEE, 2020, pp. 81--85.

\bibitem{shi2020few}
Bowen Shi, Ming Sun, Krishna~C Puvvada, Chieh-Chi Kao, Spyros Matsoukas, and
  Chao Wang,
\newblock ``Few-{S}hot {A}coustic {E}vent {D}etection via {M}eta {L}earning,''
\newblock in {\em IEEE International Conference on Acoustics, Speech and Signal
  Processing (ICASSP)}. IEEE, 2020, pp. 76--80.

\bibitem{anandfew}
Prashant Anand, Ajeet~Kumar Singh, Siddharth Srivastava, and Brejesh Lall,
\newblock ``Few {S}hot {S}peaker {R}ecognition using {D}eep {N}eural
  {N}etworks,''
\newblock {\em arXiv preprint arXiv:1904.08775}, 2019.

\bibitem{wangcentroid}
Jixuan Wang, Kuan-Chieh Wang, Marc~T Law, Frank Rudzicz, and Michael Brudno,
\newblock ``Centroid-based {D}eep {M}etric {L}earning for {S}peaker
  {R}ecognition,''
\newblock in {\em IEEE International Conference on Acoustics, Speech and Signal
  Processing (ICASSP)}. IEEE, 2019, pp. 3652--3656.

\bibitem{boneysemi}
Rinu Boney and Alexander Ilin,
\newblock ``Semi-{S}upervised {F}ew-{S}hot {L}earning with {P}rototypical
  {N}etworks,''
\newblock {\em CoRR abs/1711.10856}, 2017.

\bibitem{ayyadsemi}
Ahmed Ayyad, Nassir Navab, Mohamed Elhoseiny, and Shadi Albarqouni,
\newblock ``Semi-{S}upervised {F}ew-{S}hot {L}earning with {L}ocal and {G}lobal
  {C}onsistency,''
\newblock {\em arXiv preprint arXiv:1903.02164}, 2019.

\bibitem{lilearning}
Xinzhe Li, Qianru Sun, Yaoyao Liu, Qin Zhou, Shibao Zheng, Tat-Seng Chua, and
  Bernt Schiele,
\newblock ``Learning to {S}elf-{T}rain for {S}emi-{S}upervised {F}ew-{S}hot
  {C}lassification,''
\newblock in {\em Advances in Neural Information Processing Systems}, 2019, pp.
  10276--10286.

\bibitem{schroff2015facenet}
Florian Schroff, Dmitry Kalenichenko, and James Philbin,
\newblock ``Face{N}et: A {U}nified {E}mbedding for {F}ace {R}ecognition and
  {C}lustering,''
\newblock in {\em Proceedings of the IEEE Conference on Computer Vision and
  Pattern Recognition}, 2015, pp. 815--823.

\bibitem{renmeta}
Mengye Ren, Eleni Triantafillou, Sachin Ravi, Jake Snell, Kevin Swersky,
  Joshua~B Tenenbaum, Hugo Larochelle, and Richard~S Zemel,
\newblock ``Meta-{L}earning for {S}emi-{S}upervised {F}ew-{S}hot
  {C}lassification,''
\newblock {\em arXiv preprint arXiv:1803.00676}, 2018.

\bibitem{veauxsuperseded}
Christophe Veaux, Junichi Yamagishi, Kirsten MacDonald, et~al.,
\newblock ``Superseded-{CSTR} {VCTK} {C}orpus: {E}nglish {M}ulti-{S}peaker
  {C}orpus for {CSTR} {V}oice {C}loning {T}oolkit,''
\newblock 2016.

\bibitem{fonseca2018general}
Eduardo Fonseca, Manoj Plakal, Frederic Font, Daniel~PW Ellis, Xavier Favory,
  Jordi Pons, and Xavier Serra,
\newblock ``General-{P}urpose {T}agging of {F}reesound {A}udio with {A}udioset
  {L}abels: {T}ask {D}escription, {D}ataset, and {B}aseline,''
\newblock {\em arXiv preprint arXiv:1807.09902}, 2018.

\bibitem{vaswani2017attention}
Ashish Vaswani, Noam Shazeer, Niki Parmar, Jakob Uszkoreit, Llion Jones,
  Aidan~N Gomez, {\L}ukasz Kaiser, and Illia Polosukhin,
\newblock ``Attention is {A}ll {Y}ou {N}eed,''
\newblock in {\em Advances in Neural Information Processing Systems}, 2017, pp.
  5998--6008.

\bibitem{gemmeke2017audio}
Jort~F Gemmeke, Daniel~PW Ellis, Dylan Freedman, Aren Jansen, Wade Lawrence,
  R~Channing Moore, Manoj Plakal, and Marvin Ritter,
\newblock ``Audio {S}et: {A}n {O}ntology and {H}uman-{L}abeled {D}ataset for
  {A}udio {E}vents,''
\newblock in {\em IEEE International Conference on Acoustics, Speech and Signal
  Processing (ICASSP)}. IEEE, 2017, pp. 776--780.

\bibitem{hershey2017cnn}
Shawn Hershey, Sourish Chaudhuri, Daniel~PW Ellis, Jort~F Gemmeke, Aren Jansen,
  R~Channing Moore, Manoj Plakal, Devin Platt, Rif~A Saurous, Bryan Seybold,
  et~al.,
\newblock ``C{NN} {A}rchitectures for {L}arge-{S}cale {A}udio
  {C}lassification,''
\newblock in {\em IEEE International Conference on Acoustics, Speech and Signal
  Processing (ICASSP)}, 2017, pp. 131--135.

\bibitem{lee2013pseudo}
Dong-Hyun Lee,
\newblock ``Pseudo-{L}abel: {T}he {S}imple and {E}fficient {S}emi-{S}upervised
  {L}earning {M}ethod for {D}eep {N}eural {N}etworks,''
\newblock in {\em Workshop on Challenges in Representation Learning, ICML},
  2013, vol.~3, p.~2.

\bibitem{grandvalet2005semi}
Yves Grandvalet and Yoshua Bengio,
\newblock ``Semi-{S}upervised {L}earning by {E}ntropy {M}inimization,''
\newblock in {\em Advances in Neural Information Processing Systems}, 2005, pp.
  529--536.

\end{thebibliography}

\end{document}